\documentclass[a4]{article}

\usepackage{geometry}
\usepackage{authblk}

\usepackage{amsmath}
\usepackage{float}
\usepackage{url}
\usepackage{graphicx}

\title{\bf Functional geometry of protein-protein interaction networks}
\author[]{No\"el Malod-Dognin}
\author[]{Nata\v{s}a Pr\v{z}ulj\footnote{natasa@cs.ucl.ac.uk}}
\affil[]{Department of Computer Science, University College London, London WC1E 6BT, UK}
\date{}

\begin{document}

\maketitle

\section*{Abstract}
\noindent\textbf{Motivation:} Protein-protein interactions (PPIs) are usually modelled as networks. 
These networks have extensively been studied using graphlets, small induced subgraphs capturing the local wiring patterns around nodes in networks.
They revealed that proteins involved in similar functions tend to be similarly wired.
However, such simple models can only represent pairwise relationships and cannot fully capture the higher-order organization of protein interactions, including protein complexes.\\
\textbf{Results:} To model the multi-sale organization of these complex biological systems, we utilize simplicial complexes from computational geometry.
The question is how to mine these new representations of PPI networks to reveal additional biological information.
To address this, we define {\em simplets}, a generalization of graphlets to simplicial complexes.
By using simplets, we define a sensitive measure of similarity between simplicial complex network representations that allows for clustering them according to their data types better than clustering them by using other state-of-the-art measures, e.g., spectral distance, or facet distribution distance.\\
We model human and baker's yeast PPI networks as simplicial complexes that capture PPIs and protein complexes as simplices.
On these models, we show that our newly introduced simplet-based methods cluster proteins by function better than the clustering methods that use the standard PPI networks, uncovering the new underlying functional organization of the cell.
We demonstrate the existence of the functional geometry in the PPI data and the superiority of our simplet-based methods to effectively mine for new biological information hidden in the complexity of the higher order organization of PPI networks.\\

\section{Introduction}
\subsection{Motivation}
Genome is the blueprint of a cell. DNA regions called genes are transcribed into messenger RNAs that are translated into proteins.
These proteins interact with each other and with other molecules to perform their biological functions.
Deciphering the patterns of molecular interactions (also called topology) is fundamental to understanding the functioning of the cell \cite{ryan2013}.
In system biology, molecular interactions are modeled as various molecular interaction networks, in which nodes represent molecules and edges connect molecules that interact in some way.
Examples include the well-known protein-protein interaction (PPI) networks in which nodes represent proteins and edges connect proteins that can physically bind.

Because exact comparison between networks has long been known to be computationally intractable \cite{cook1971}, the topological analyses of biological networks use approximate comparisons (heuristics), commonly called network properties, such as the degree distribution, to approximately say whether the structures of networks are similar \cite{newman2010}. 
Advanced network properties that utilize graphlets (small induced subgraphs) \cite{przulj2004} have been successfully used to mine biological network datasets.
Graphlet-based properties include measures of topological similarities between nodes and between networks \cite{przulj2004,przulj2007,yaveroglu2014}, as well as between protein 3D structures represented by networks \cite{malod2014gr, faisal2017}.
In particular, graphlets have been used to characterize and compare the local wiring patterns around nodes in a PPI network \cite{milenkovic2008}, which revealed that molecules involved in similar functions tend to be similarly wired \cite{davis2015}.
These topological similarities between nodes have also been used to to guide the node mapping process of network alignment methods \cite{kuchaiev2010,kuchaiev2011,malod2015,vijayan2015}, which allowed for transferring of biological annotation between nodes in different networks of well-studied species to less studied ones.

Despite significant progress, these simple network (also called graph) models of molecular interaction data can only represent pairwise relationships and cannot fully capture the higher organization of molecular interactions, such as protein complexes and biological pathways \cite{estrada2005}.
Hence, we need to model these data by using new mathematical formalisms capable of capturing their multi-scale organization.
Furthermore, we need to design new algorithms capable of extracting new biological information hidden in the wiring patterns of the molecular interaction data modeled by using these mathematical formalisms.
This paper addresses these issues.

\subsection{Simplicial complexes basics}
\begin{figure}
	\begin{centering}
	\includegraphics[width=6cm]{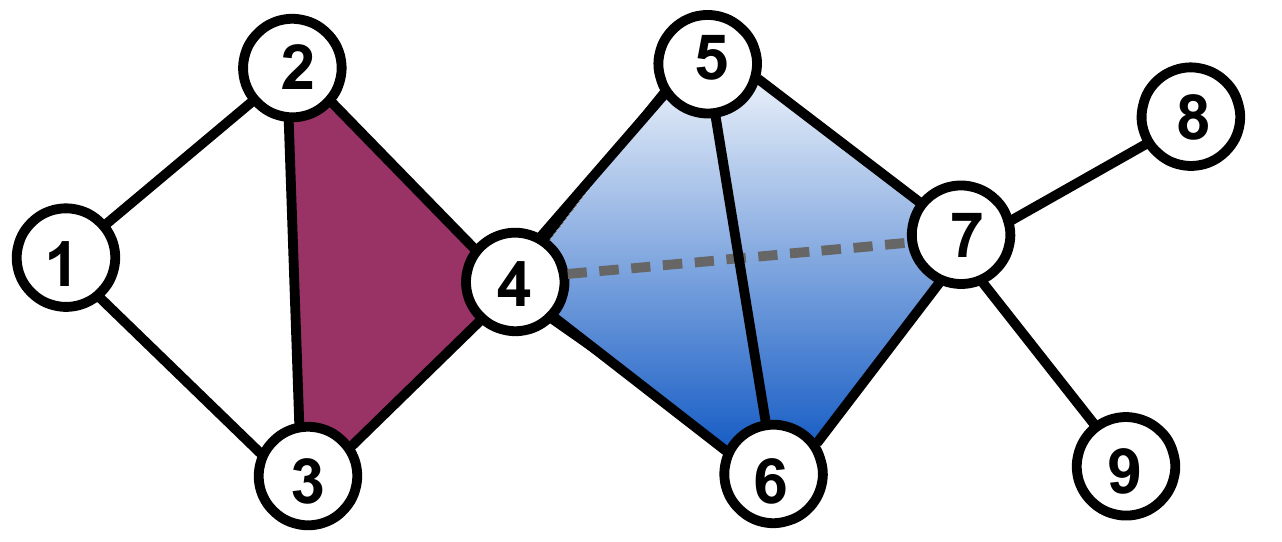}
	\caption{{\bf Illustration of a 3-dimensional simplicial complex.} In the presented simplicial complex, nodes 1, 2 and 3 are only connected by 1-dimensional simplices (edges, in black). Nodes 2, 3 and 4 are connected by a 2-dimensional simplex (triangle, in magenta). Nodes 4, 5 , 6 and 7 are connected by a 3-dimensional simplex (tetrahedron, in blue).
 {\label{fig:complex}}}
	\end{centering}
\end{figure}

A candidate model for capturing higher-order molecular organization is a simplicial complex \cite{munkres1984}.
A {\em simplicial complex} is  a set of {\em simplices}, where a 0-dimensional simplex is a node, a 1-dimensional simplex is an edge, a 2-dimensional simplex is a triangle, a 3-dimensional simplex is a tetrahedron and their $n$-dimensional counterparts (illustrated in Figure \ref{fig:complex}).
The dimension of a simplicial complex is the largest dimension of its simplices.

The $(n\text{-}1)$-dimensional sub-simplices of an $n$-dimensional simplex are called its {\em faces} (e.g., a triangle has three faces, the three edges).
A simplicial complex, $K$, is required to satisfy two conditions:
\begin{itemize}
\item For any simplex $\delta \in K$, any face $\delta'$ of $\delta$ is also in $K$.
\item For any two simplices, $\delta_1, \delta_2 \in K$, $\delta_1 \bigcap \delta_2$ is either $\emptyset$, or a face of both  $\delta_1$ and $\delta_2$.
\end{itemize}
In a simplicial complex, a {\em facet} is a simplex that is not a face of any higher dimensional simplex. Because of this property, a simplicial complex can be summarized by its set of facets.

Note that a network is a 1-dimensional simplicial complex and thus, our proposed methodology is directly applicable to both traditional networks and the higher dimensional simplicial complexes.

While simple network statistics, such as degrees, shortest paths and centralities, have been generalized to simplicial complexes \cite{estrada2018}, the lack of more advanced statistics capturing the geometry of simplicial complexes limits their usage in practical applications

\subsection{Contributions}

To comprehensively capture the multi-scale organization of complex molecular networks, we propose to model them by using simplicial complexes.
To extract the information hidden in the geometric patterns of these models, we generalize graphlets to simplicial complexes, which we call {\em simplets}.
On large scale real-world and synthetic simplicial complexes, we show that simplets can be used to define a sensitive measure of geometric similarity between simplicial complexes.
Then, on simplicial complexes capturing the protein interactomes of human and yeast, we show that simplets can be used to relate the local geometry around proteins in simplicial complexes with their biological functions.
Comparison between 1-dimensional protein-protein interaction networks and the higher-dimensional simplicial complex representations of the interactomes formed by protein interactions and protein complexes shows that higher-order modeling enabled by simplicial complexes allows for capturing more biological information, which can efficiently be mined with our proposed simplets.

\section{Methods}

\subsection{Datasets and their simplicial complex representations}

\subsubsection{Yeast and human protein interactomes}\label{sec:models}
From BioGRID (v. 3.4.156)\cite{chatr2017}, we collected the experimentally validated protein-protein interaction (PPI) networks of human (H. sapiens) and of yeast (S. cerevisiae).
From CORUM \cite{ruepp2010}, we collected collected on the $2^{nd}$ of July, 2017) the experimentally validated protein complexes of human, and from CYC2008 (v.2.0) \cite{pu2009} the experimentally validated protein complexes of yeast.
We consider three different models of an organism's interactome.
\begin{itemize}
	\item {\bf The 1-dimensional PPI model}: it is the usual PPI network, in which proteins (nodes) are connected by an edge if they can physically bind. Recall that a network is a 1-dimensional simplicial complexes on which our new simplet methodologies can be applied and are equivalent to the standard graphlet methodologies.
	\item {\bf The higher-dimensional simplicial complex (SC) model}: starting from the PPI network, we additionally connect by simplicies all the proteins that belong to common complexes. I.e., the proteins belonging to a $k$-protein complex are connected by a $(k\text{-} 1)$ dimensional simplex.
\end{itemize}

For human, the PPI network (1D PPI model) has 16,100 nodes and 212,319 edges.
When unifying the lower dimensional protein-protein interaction data and the higher order protein complex data as described above,
the resulting SC model is a 140-dimensional simplicial complex having 16,140 nodes (with 40 proteins being part of proteins complexes but not having any reported protein-protein interaction) and 205,192 facets.
For yeast, the 1D PPI model has 5,842 nodes and 80,900 edges.
When unifying the lower dimensional protein-protein interaction data and the higher order protein complex data as described above,
the SC model is a 80-dimensional simplicial complex having 5,842 nodes and 76,790 facets.

\subsubsection{Other real-world datasets}\label{sec:randoms}
We collected real-world higher-dimensional datasets from biology and beyond.
\begin{itemize}
\item {\bf 1,569 simplicial complexes of protein 3D structures:} Proteins are linear arrangements of amino-acids that in the aqueous environment of the cell fold and acquire specific three-dimensional (3D) shapes called tertiary structures. We collected from Astral-40 (SCOPe v.2.06) \cite{fox2013} the 3D structures of 1,569 protein domains that are at-least 100 amino-acid long. Each protein domain is modeled as a simplicial complex in which simplices connect together all the amino-acids (nodes) that are less than 7.5 \AA ~apart (as measured by the distances between their $\alpha$-carbons).\\

\item {\bf 132 simplicial complexes of publication authorships:} From the preprint repository arXiv, we collected all the scientific publications in the ``computer science'' category over eleven years from 2007 to 2017. For each  month, we model the scientific collaborations as a simplicial complex in which simplices are formed by all scientists (nodes) that co-authored a scientific publication.\\

\item {\bf 60 simplicial complexes of genes' biological annotations:} The biological functions of genes are described by various ontology terms. We collected pathway annotations from Reactome database (v.63) \cite{fabregat2017}, as well as the experimentally validated Gene Ontology (GO)\cite{ashburner2000} annotations from NCBI's entrez web-server (collected in February 2018). For GO, we consider biological process, molecular function, and cellular component annotations separately. For each annotation set, we model the functional annotations of the genes of a given species as a simplicial complex in which simplices are formed by all genes (nodes) that have a common annotation term (restricted to terms annotating up-to 50 genes for computational complexity issues). We only considered simplicial complexes having more than 100 nodes. Following this procedure, we generated 18 pathway simplicial complexes, 13 biological process simplicial complexes, 14 molecular function simplicial complexes and 15 cellular component simplicial complexes.\\
	
\item {\bf14 simplicial complexes of protein-protein interactions:} We collected the experimentally validated protein-protein interactions (PPIs) from BioGRID database (v. 3.4.156)\cite{chatr2017}. These PPIs are first modeled as networks in which proteins (nodes) are connected by edges if they can interact. The corresponding networks are converted into so-called {\em clique complexes}, by creating a simplex between all nodes belonging to a maximal clique in the network.
\end{itemize}

\subsubsection{Random simplicial complexes}\label{sec:reals}
To test our methods, we considered randomly generated simplicial complexes, which we generate according to eight random models.
The first four models are based on randomly generated graphs, which are converted into so-called {\em clique complexes}, in which simplices connect nodes that belong to a clique in the graph:
\begin{itemize}
\item A {\em random clique complex} (RCC) is the clique complex of an Erd\"{o}s-R\`{e}nyi random graph \cite{erdos1959}. An Erd\"{o}s-R\`{e}nyi graph is generated by fixing the number of nodes (detailed below) in the graph, and then by adding edges between uniformly randomly chosen pairs of nodes until a given edge density is reached (also detailed below).
\item A {\em Vietoris-Rips complex} (VRC) \cite{hausmann1995} is the clique complex of a geometric random graph \cite{penrose2003}. A geometric random graph represents the proximity relationship between uniformly randomly distributed points in a $d$-dimensional space. We generate geometric graphs by uniformly randomly distributing the desired number of nodes (points) in a 3-dimensional unit cube. Then, two nodes are connected by an edge if the Euclidean distance between the corresponding points is smaller than a distance threshold $r$. The distance threshold is chosen to obtain the desired edge density.
\item A {\em scale-free complex} (SFC) is the clique complex of a Barab\`{a}si-Albert scale-free graph \cite{barabasi1999}. The scale-free graph model constructed by preferential attachment generates graphs based on the ``rich-get-richer'' principle and are characterized by power-law degree distributions. We create a scale-free graph using an iterative process, in which the graph is grown by attaching new nodes each with $m$ edges that are preferentially attached to the existing nodes with high degree ($m$ is chosen to obtain the desired edge density).
\item A {\em Watts-Strogatz complex} (WCS) is the clique complex of a small-world graph \cite{watts1998}. Small-world graphs are characterized by short average path lengths and high clustering. We created a small word graph by constructing a regular ring lattice of $n$ nodes and by connecting each node to its $k$ neighbours, $k/2$ on each side ($k$ is chosen to obtain the desired edge density). Then we uniformly randomly rewire 5\% of the edges.
\end{itemize}
The four other models are extensions of the {\em Linial-Meshulam} model \cite{linial2006,meshulam2009}, which originally consists in randomly connecting nodes with $k$-dimensional facets. We extended this model to randomly connect nodes with facets while following the facet distribution of an input simplicial complex.
In this way, we can create Linial-Meshulam variant of the four clique complex-based models presented above: 
\begin{itemize}
\item A {\em  Linial-Meshulam random clique complex} (LM- RCC) is a Linial-Meshulam complex that follows the facet distribution of an input random clique complex.
\item A {\em  Linial-Meshulam Vietoris-Rips complex} (LM- VRC) is a Linial-Meshulam complex that follows the facet distribution of an input Vietoris-Rips complex.
\item A {\em  Linial-Meshulam scale-free complex} (LM-SFC) is a Linial-Meshulam complex that follows the facet distribution of an input scale-free complex.
\item A {\em  Linial-Meshulam Watts-Strogatz complex} (LM- WSC) is a Linial-Meshulam complex that follows the facet distribution of an input Watts-Strogatz complex.
\end{itemize}

For each model we choose three node sizes, 1,000, 2,000, and 3,000 nodes, and three edge densities, 0.5\%, 0.75\% and 1\%. We generated 25 random simplicial complexes for each model and each of these node sizes and edge densities. Hence, in total, we generated $8\times 3\times 3\times 25 =$ 1,800 random simplicial complexes.
We chose these node sizes and edge densities to roughtly mimic the sizes and densities of real-world data detailed above.

\subsection{Capturing the local geometry around nodes in a simplicial complex with simplets}\label{sec:simplets}
\begin{figure}[H]
	\begin{centering}
	\includegraphics[width=8cm]{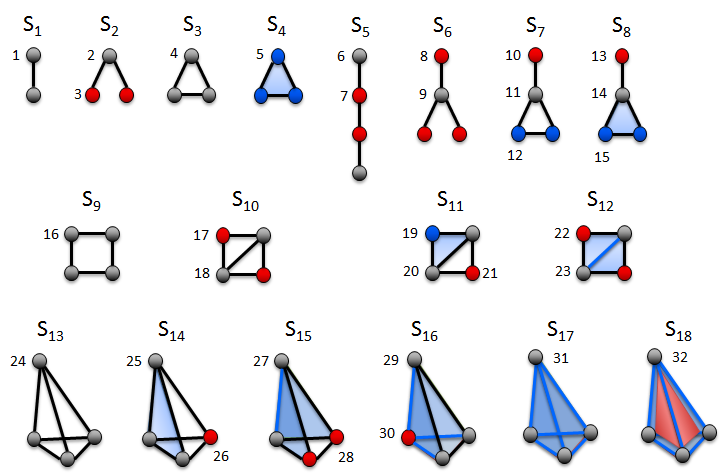}
	\caption{{\bf Illustration of 2- to 4-nodes simplets.} The 18 2- to 4-nodes simplets are denoted by $S_1$ to $S_{18}$.
	Within each simplet, geometrically interchangeable nodes, belonging to the same orbit, have the same color.
	These simplets have 32 different orbits, denoted from 1 to 32.
	Note that simplets  $S_4$, $S_8$, $S_{11}$ and $S_{14}$ have only one 2D face (triangle, in blue), while $S_{12}$ and $S_{15}$ have two triangles, $S_{16}$ has 3 triangles and $S_{17}$ has four triangles. $S_{18}$ has four triangles and one 3D face (tetrahedron, in red).
 {\label{fig:simplets}}}
	\end{centering}
\end{figure}

{\em Simplets} are small, connected, non-isomorphic, induced simplicial complexes of a larger simplicial complex. Figure \ref{fig:simplets} shows the eighteen 2- to 4-node simplets (denoted by $S_1$ to $S_{18}$).
Within each simplet, because of symmetries, some nodes can have identical geometries.
Analogous to automorphism orbits in graphlets \cite{przulj2007}, we say that such nodes belong to a common {\em simplet orbit group}, or {\em orbit} for brevity. Figure \ref{fig:simplets} shows the thirty-two orbits of the 2- to 4-node simplets (denoted from 1 to 32).
Similar to graphlets, we use simplets to generalize the notion of the node degree:
the $i^{th}$ {\em simplet degree} of node $v$, denoted by $v_i$, is the number of times node $v$ touches a simplet at orbit $i$.

We define the {\em simplet degree vector} (SDV) of a node as the 32 dimensional vector containing the simplet degrees of the node in the simplicial complex as its coordinates.
Hence, the SDV of a node describes the local geometry around the node in the simplicial complex and comparing the SDVs of two nodes provides a measure of local geometric similarity between them.

We define the {\it SDV similarity} between two nodes as an extension of the graphlet degree similarity \cite{milenkovic2008}. It is computed as follows.
The distance, $D_{i}(u,v)$, between the $i^{th}$ simplet orbits of nodes $u$ and $v$ is defined as:
\begin{equation}
	D_i(u,v) = w_i \times \frac{|log(u_i + 1) - log(v_i + 1)|}{log(max\{u_i, v_i\} + 2)},
\end{equation}
where $w_{i}$ is the weight of orbit $i$ that accounts for dependencies between orbits.
Weight,  $w_{i}$, is computed as $\displaystyle w_i = 1 - \frac{\log (o_i)}{\log (32)}$, 
where $o_i$ is the number of orbits that orbit $i$ depends on, including itself.
For instance, the count of orbit 2 (the middle of a three node path) of a node depends on its count of orbit 0 (i.e. its node degree) and on itself, so $o_2 = 2$.
For orbit 9, $o_{9}$ = 3, since it is affected by orbits 0, 2, and itself. The values of $o_i$ for all 2- to 4-nodes simplet orbits are listed in Table \ref{Tab:01}.

\begin{table}
\begin{centering}
\begin{tabular}{c c}
Orbit, $i$ & Weight, $o_i$\\
\hline
1 & 1 \\
2, 3, 4, 5 & 3 \\
6, 8, 9, 10, 13, 24, 26, 30, 31, 32 & 3 \\
7, 11, 12, 14, 15, 16, 17, 18, 19, 21, 22, 23, 25, 27, 28, 29 & 4\\
20 & 5\\
\hline
\end{tabular}
\caption{The orbit weights.\label{Tab:01}}
\end{centering}
\end{table}

Finally, the SDV similarity, $S(u,v)$, between nodes $u$ and $v$ is defined as:
\begin{equation}
	S(u,v) = 1 - \frac{\sum_{i | (u_i \neq 0) \text{ or } (v_i \neq 0)}D_i(u,v)}{\sum_{i | (u_i \neq 0) \text{ or } (v_i \neq 0)}w_i}.
\end{equation}
$S(u,v)$ is in (0, 1], where similarity 1 means that the  SDVs of nodes $u$ and $v$ are identical.\\

\subsection{Capturing the global geometry of a simplicial complex with simplets}\label{sec:distances}
To the best of our knowledge, researchers from computational geometry have not considered the problem of comparing two simplicial complexes. However, the comparison of biological networks is a foundational problem of system biology.
Instead, computational geometry focus on the comparison of two spaces, each represented by a collection of simplicial complexes, e.g. \cite{collins2004}.
Thus, we build upon network analysis and extend graphlet and non-graphlet based network distance measures to directly compare simplicial complexes as follows.

\subsubsection{Simplet correlation distance}\label{sec:scd}
Simplets are like Lego pieces that assemble with each other to build larger simplicial complexes. We exploit this property to summarize the complex structures of simplicial complexes and to compare them, by generalizing Graphlet Correlation Distance \cite{yaveroglu2014}, which is a sensitive measure of topological similarity between networks.

Analogous to graphlets, the statistics of different simplet orbits are not independent of each other.
The reason behind this is the fact that smaller simplets are induced sub-simplicial complexes of larger simplets.
For 2- to 4-node simplets, there are four non-redundant dependency equations between the simplet degrees of a given node $u$:
\begin{equation}
{{u_1}\choose{2}} = u_2 + u_4 + u_5,
\end{equation}
\begin{equation}
{{u_2}\choose{1}}{{u_1-2}\choose{1}} = 3u_9 + 2u_{11} + 2u_{14} + u_{18} + u_{20} +u_{23},
\end{equation}
\begin{equation}
{{u_3}\choose{1}}{{u_1-1}\choose{1}} = \begin{array}{l}u_7 + u_{12} + u_{15} + 2u_{16} +2u_{17}\\ + 2u_{19} + 2u_{21} + 2u_{22}\end{array},
\end{equation}
\begin{equation}
{{u_1}\choose{3}} = \begin{array}{l}u_9 + u_{11} + u_{14} + u_{18} + u_{20} + u_{23} + u_{24} + u_{25}\\ + u_{26} + u_{27} + u_{28} + u_{29} + u_{30} + u_{31} + u_{32}\end{array}.
\end{equation}
We used these equations to assess the correctness of our exhaustive simplet counter.

\begin{figure}
	\begin{centering}
	\begin{tabular}{c c}
	RCC & VRC \\
	\includegraphics[width=4cm]{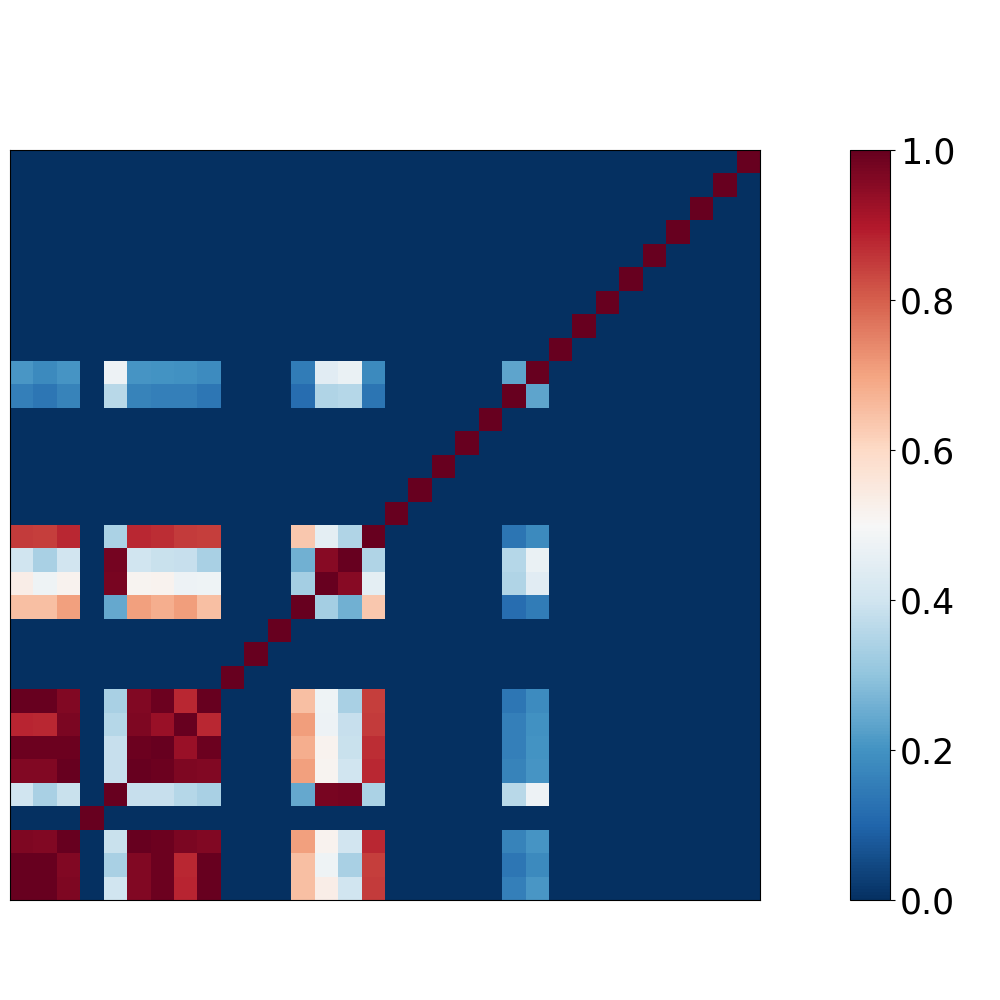} &
	\includegraphics[width=4cm]{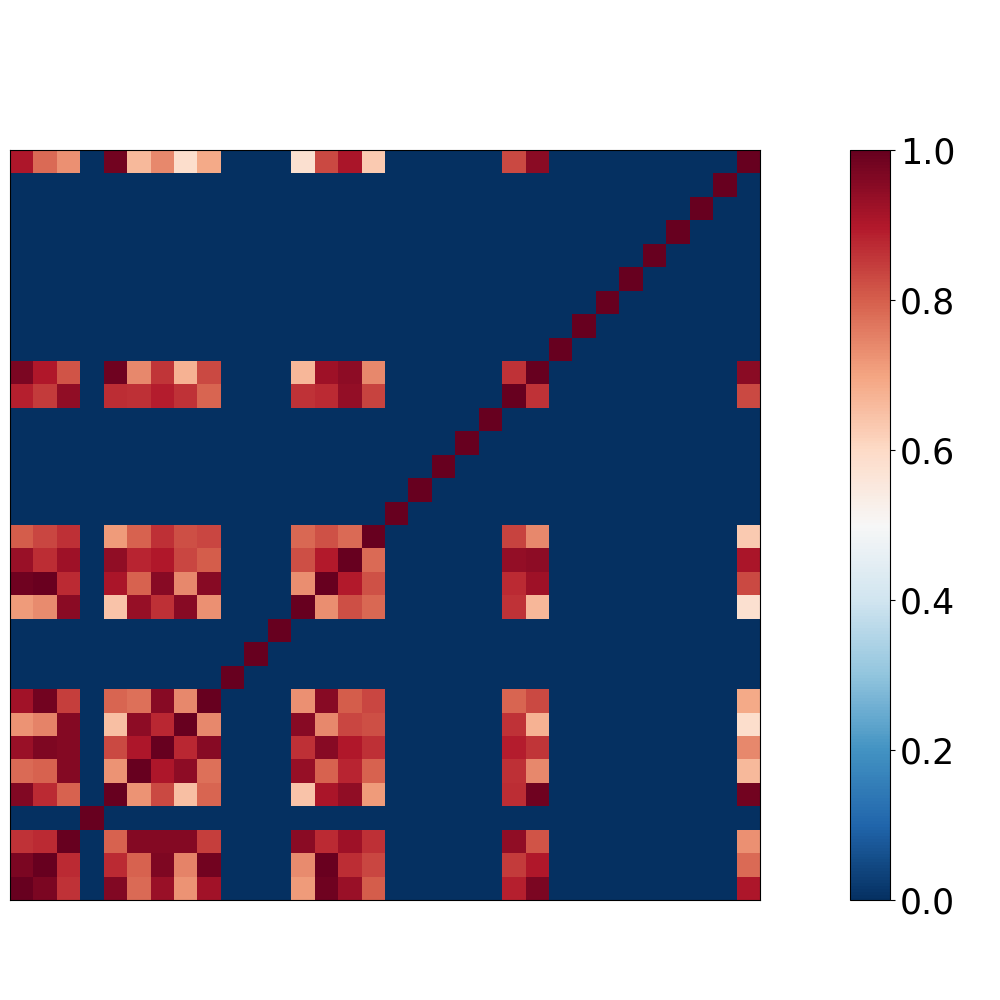} \\
	~\\
	LM-RCC & LM-VRC \\
	\includegraphics[width=4cm]{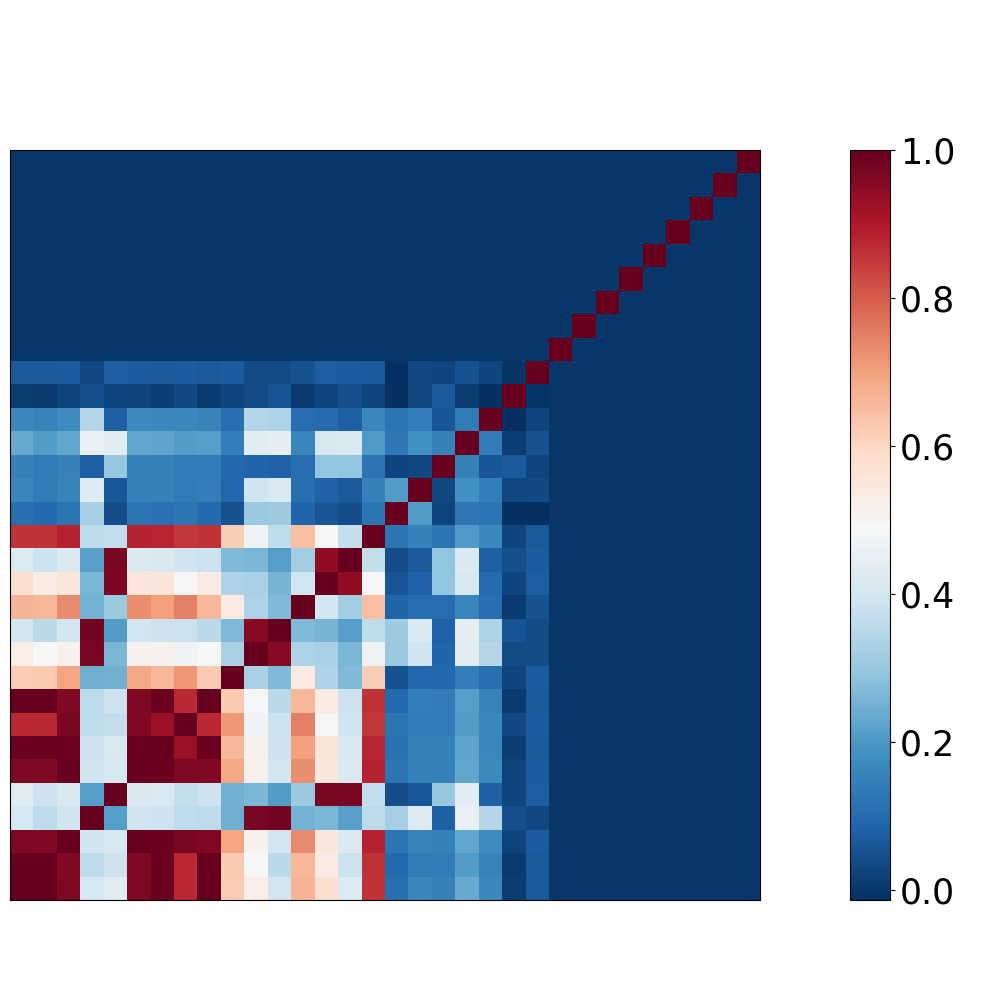} &
	\includegraphics[width=4cm]{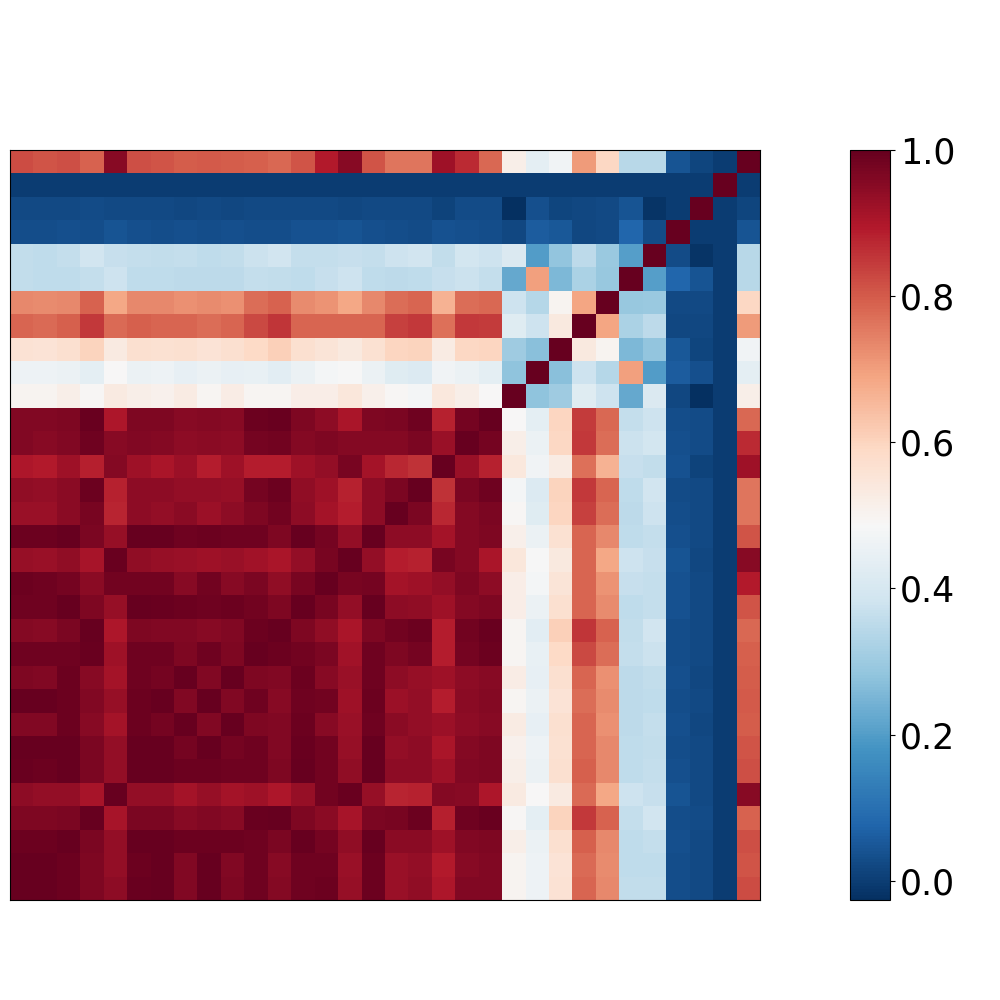} \\
	\end{tabular}
	\caption{{\bf SCMs of sample simplicial complexes from four random simplicial complex models.} The four simplicial complexes from RCC, VRC, LM-RCC, and LM-VRC models have been generated with 2,000 nodes and the edge density of 0.75\%.
	Note that SCMs of all networks of this size and density coming from a particular model look similar. 
	Hence, these four SCMs are representative of these models at these sizes and densities.
 {\label{fig:scms}}}
	\end{centering}
\end{figure}

In addition to these redundancies there also exist dependencies between simplets, which are dataset dependent.
We use these dataset dependent simplet orbit dependencies to characterize the global geometry of simplicial complexes.
We capture the dependencies between simplet orbits by the simplicial complex's {\em Simplet Correlation Matrix} (SCM), which we define as follows.
We construct a matrix whose rows are the simplet degree vectors of all nodes of the simplicial complex.
We calculate the Spearman's correlation between each two pairs of columns in the resulting matrix, i.e., correlations between the orbits overl all nodes of the simplicial complex.
We present these correlations in a $32 \times 32$  dimensional Simplet Correlation Matrix (SCM): it is symmetric and contains Spearman's correlation values in [-1,1] range.
As presented in Figure \ref{fig:scms}, the SCMs of simplicial complexes from different random simplicial complex models are indeed very different. We exploit these differences in SCMs to compare simplicial complexes.

We define the {\em Simplet Correlation Distance} (SCD) to measure the distance between two simplicial complexes, $K_1$ and $K_2$,  by the Euclidean distance between the upper-triangles of their SCMs:
\begin{equation}
SCD(K_1, K_2) = \sqrt{ \sum_{i=1}^{32} \sum_{j=i+1}^{32} (SCM_{K_1}[i][j] - SCM_{K_2}[i][j])^2},
\end{equation}
where $SCM_{K_1}[i][j]$ is the $(i,j)^{th}$ entry in the SCM of $K_1$ (similar for $K_2$).
The ability of SCD to group together simplicial complexes according to their underlying models is demonstrated in section \ref{sec:res_clustering}.

\subsubsection{Facet distribution distance}
In analogy to degree distribution and graphlet degree distribution \cite{przulj2007}, we define the measure of connectivity of a $k$-dimensional simplicial complex, $K$, as the distribution of its facets,  $d_{K}$: it is a $k$-dimensional {\em facet distribution vector} whose $i^{th}$ entry is the percentage of the facets in $K$ having dimension $i$.
The {\em Facet Distribution Distance} (FDD) measures the distance between two simplicial complexes, $K_1$ and $K_2$, by the Euclidean distance between their facet distribution vectors, $d_{K_1}$ and $d_{K_2}$:

\begin{equation}
FDD(K_1, K_2) = \sqrt{\sum_{i}{(d_{K_1}[i] -d_{K_2}[i])^2}}.
\label{eq:specDist}
\end{equation}

\subsubsection{Spectral distance}
Spectral theory captures the topology of networks and simplicial complexes by using the eigenvalues and eigenvectors of matrices representing them, such as the adjacency matrix, or Laplacian matrix \cite{wilson08}.
Let $H$ be the incidence matrix of a simplicial complex, $K$, having $n$ nodes and $f$ facets: $H$ is a $n\times f$ matrix in which entry $H[i][j] = 1$ if node $i$ is in facet $j$, and 0 otherwise.
The corresponding degree matrix, $D$, is a $n \times n$ diagonal matrix in which entry $D[i][i]$ is the number of facets containing node $i$.
The adjacency matrix, $A$, of a simplicial complex is:
\begin{equation}
	A = HH^T - D,
\end{equation}
where $H^T$ is the transpose of $H$ \cite{zhou2007}.
The corresponding Laplacian matrix, $L$, is defined as \cite{zhou2007}:
\begin{equation}
L =\frac{1}{2} D^{-1/2}AD^{-1/2}.
\end{equation}

The eigen-decomposition of the Laplacian matrix, $L$, of simplicial complex, $K$, is $L = \phi\lambda_K\phi^{T}$, where $\lambda_K = diag(\lambda_{K}^1, \lambda_{K}^2, ..., \lambda_{K}^n)$ is the diagonal matrix with the ordered eigen-values, $\lambda_{K}^i$ as elements and $\phi = (\phi_{1}|\phi_{2}|...|\phi_{n})$ is the matrix with the ordered eigen-vectors as columns.
The spectrum of simplicial complex, $K$, is the set of its eigen-values $S_K = \{\lambda_{K}^1, \lambda_{K}^2, ..., \lambda_{K}^n\}$, which are reordered so that $\lambda_{K}^1 \geq \lambda_{K}^2 \geq ... \geq\lambda_{K}^n$.

We define the {\em spectral distance} (SD) between two simplicial complexes, $K_1$ and $K_2$, as the Euclidean distance between their spectra \cite{wilson08}:
\begin{equation}
SD(K_1, K_2) = \sqrt{\sum_{i}{(\lambda_{K_1}^i - \lambda_{K_2}^i)^2}}.
\label{eq:specDist}
\end{equation}
When the two spectra are of different sizes, $0$ valued eigenvalues are added at the end of the smaller spectrum.

\section{Results and discussion}

\subsection{Comparing simplicial complexes}\label{sec:res_clustering}
\begin{figure}
	\begin{centering}
	\includegraphics[width=8cm]{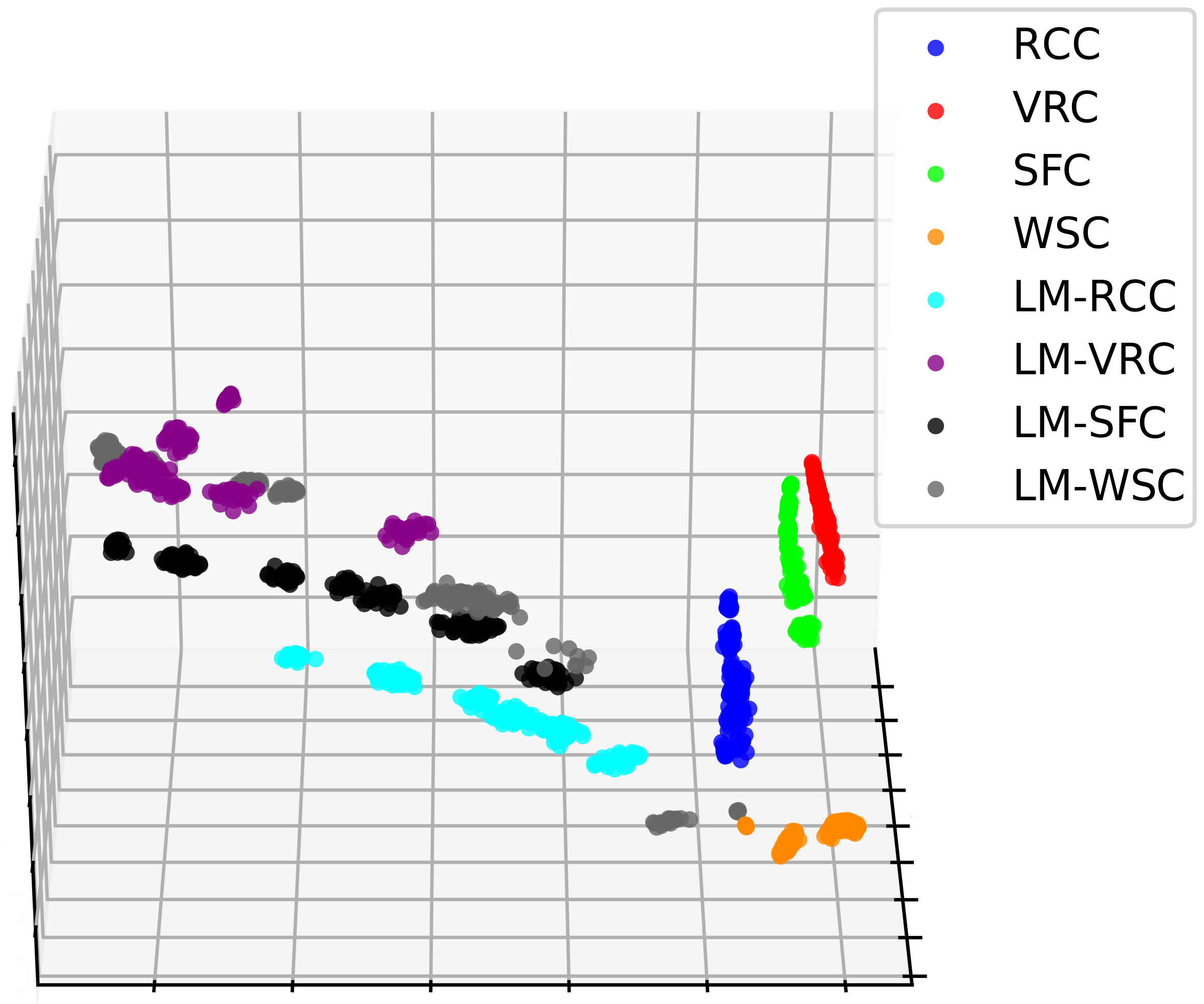}
	\caption{{\bf Illustration of MDS-based embedding of simplicial complexes from eight random models.} The randomly generated simplicial complexes (color-coded) are embedded into 3D space according to their pairwise SCD distances using multi-dimensional scaling (MDS). The eight models and simplicial complex sizes and densities are described in Section \ref{sec:randoms}.
	As described in Section  \ref{sec:randoms}, 25 simplicial complxes are generated for each model and each of its sizes and densities. The grouping of the same colored nodes correspond to simplicial complexes from the same model, but of different sizes and densities.
 {\label{fig:scd_random1}}}
	\end{centering}
\end{figure}

We visually inspect how well our simplet correlation distance (SCD, presented in section \ref{sec:scd}) groups simplicial complexes of the same type, by embedding simplicial complexes as points in 3D space according to their pairwise SCDs by using multi-dimensional scaling (MDS, \cite{borg2005}).
As presented in Figure \ref{fig:scd_random1}, when the simplicial complexes are embedded into 3D space by using multi-dimensional scaling, they visually form clusters.

We formally assess the clustering ability of SCD and of other distances between simplicial complexes by using the standard Precision-Recall and Receiver Operating Characteristic (ROC) curves analyses \cite{fawcett2006}.
For small increments of parameter $\epsilon \geq 0$, if the distance between two simplicial complexes is smaller than $\epsilon$, then the pair of simplicial complexes is declared to be similar (belong to the same cluster). For each $\epsilon$, four values are computed:
\begin{itemize}
\item The true positives (TP) are the numbers of correctly clustered pairs of simplicial complexes (grouping together simplicial complexes from the same model).
\item The true negatives (TN) are the numbers of correctly non-clustered pairs of simplicial complexes (not grouping together simplicial complexes from different models).
\item The false positives (FP) are the numbers of incorrectly clustered pairs of simplicial complexes (grouping together simplicial complexes from different models).
\item the false negatives (FN) are the numbers of incorrectly non-clustered pairs of simplicial complexes (not grouping together simplicial complexes from the same model).
\end{itemize}
In Precision-Recall curves, for each $\epsilon$, the precision ($Pr=\frac{TP}{TP+FP}$) is plotted against the recall, also-called true-positive rate ($Re=\frac{TP}{TP+FN}$). The quality of the grouping is measured with the area under the Precision-Recall curve, which is the {\em average precision} (AP) of the distance measure.
In ROC curves, for each $\epsilon$, the true positive rate (another name for recall) is plotted against the false positive rate ($FPR=\frac{FP}{FP+TN}$). The quality of the grouping is measured with the area under the ROC curve (AUC), which can be interpreted as the probability that a randomly chosen pair of simplicial complexes coming from the same model will have a distance smaller than a randomly chosen pair of simplicial complexes coming from different models.

\begin{figure}
	\begin{centering}
	\includegraphics[width=8cm]{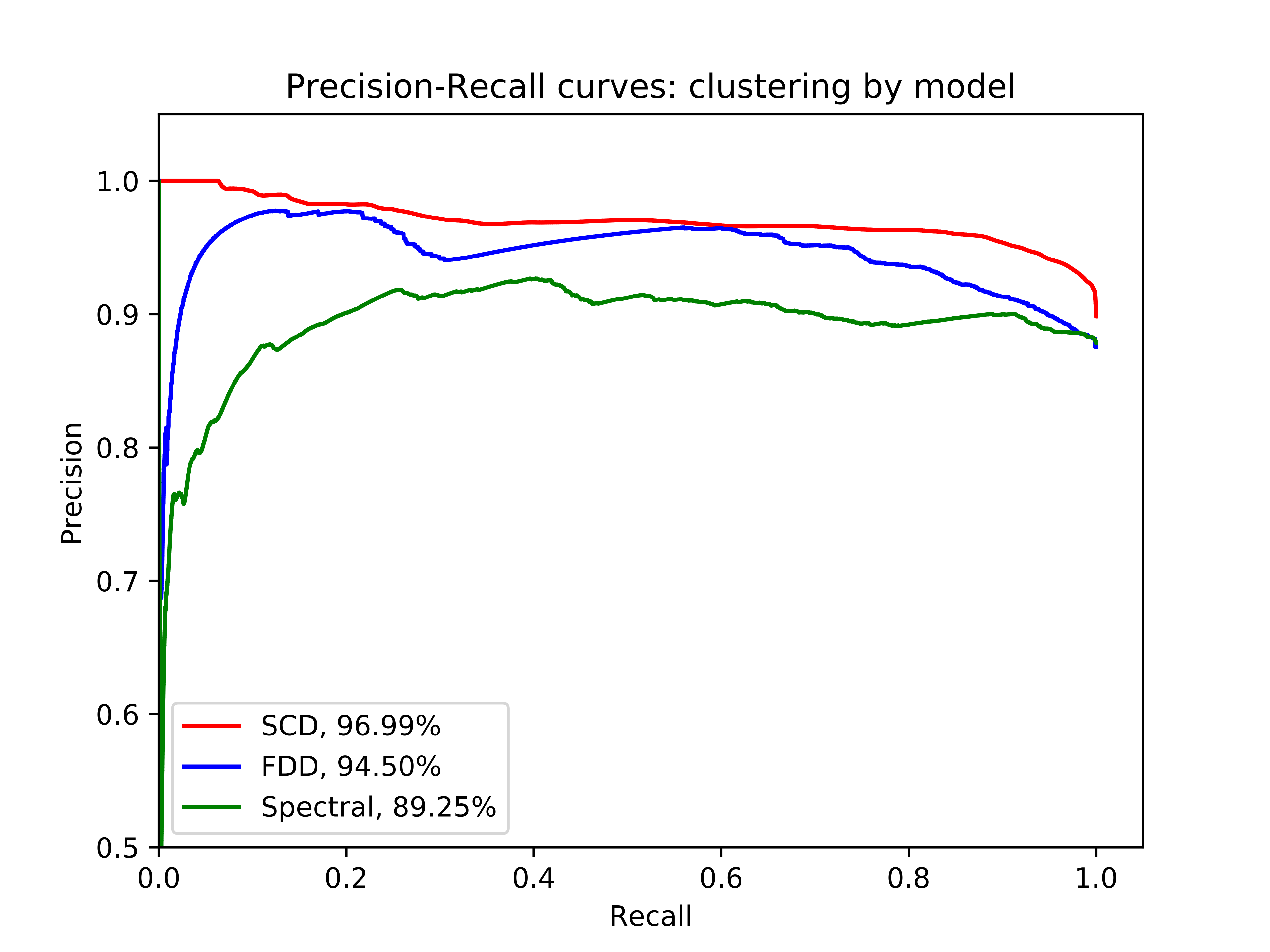}
	\caption{{\bf Clustering randomly generated simplicial complexes.}The Precision-Recall curves that are achieved when using the three distance measures (color coded, simplet correlation distance in red, facet distribution distance in blue, spectral distance in green) to cluster together the 1,800 randomly generated simplicial complexes into the models that generated them.
 {\label{fig:scd_random2}}}
	\end{centering}
\end{figure}

First, we consider our 1,800 randomly generated simplicial complexes (Section \ref{sec:randoms}), and all 1,619,100 pairs of these 1,800 simplicial complexes, to measure the ability of SCD to group together the simplicial complexes from the same model. 
The precision recall curves presented in Figure \ref{fig:scd_random2} confirm our visual illustration of the ability of SCD to classify simplicial complexes.
To assess the performance of SCD, we first apply it to synthetic data. In particular, we apply the three distance measures described in Section \ref{sec:distances} to the 1,800 model simplicial complexes described in Section \ref{sec:randoms}.
We find that SCD achieves the highest classification performance with average precision (AP) of 96.99\% and an AUC of 86.03\%.
It is followed by the facet distribution distance (AP of 94.50\% and AUC of 76.95\%) and by the spectral distance (AP of 89.25\% and AUC of 61.26\%).
Furthermore, on the easier task of grouping together simplicial complexes that are generated from the same models and the same node sizes and edge densities, SCD achieves an almost perfect clusterings having average precision of 99.99\% and AUC of 99.17\%. It is followed by facet distribution distance (AP of 99.98\% and AUC of 98.50\%) and by spectral distance (AP of 99.98\% and AUC of 98.23\%). 

\begin{figure}
	\begin{centering}
	\includegraphics[width=8cm]{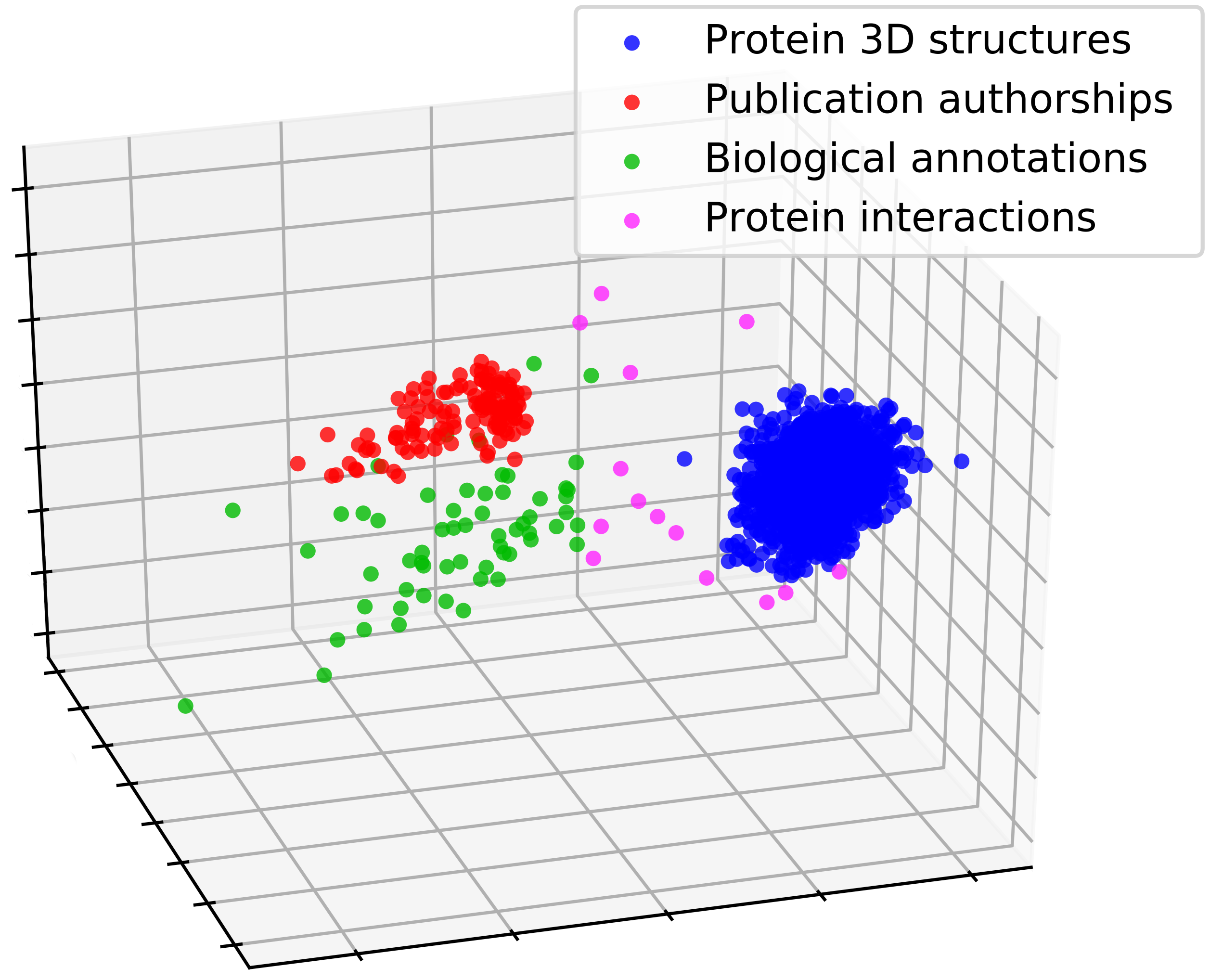}
	\caption{{\bf Illustration of MDS-based embedding of real-world simplicial complexes based on their SCDs.} The real-world simplicial complexes (color-coded) are embedded into 3D space according to their pairwise SCD distances using multi-dimensional scaling.
 {\label{fig:scd_real1}}}
	\end{centering}
\end{figure}

\begin{figure}
	\begin{centering}	
	 \includegraphics[width=8cm]{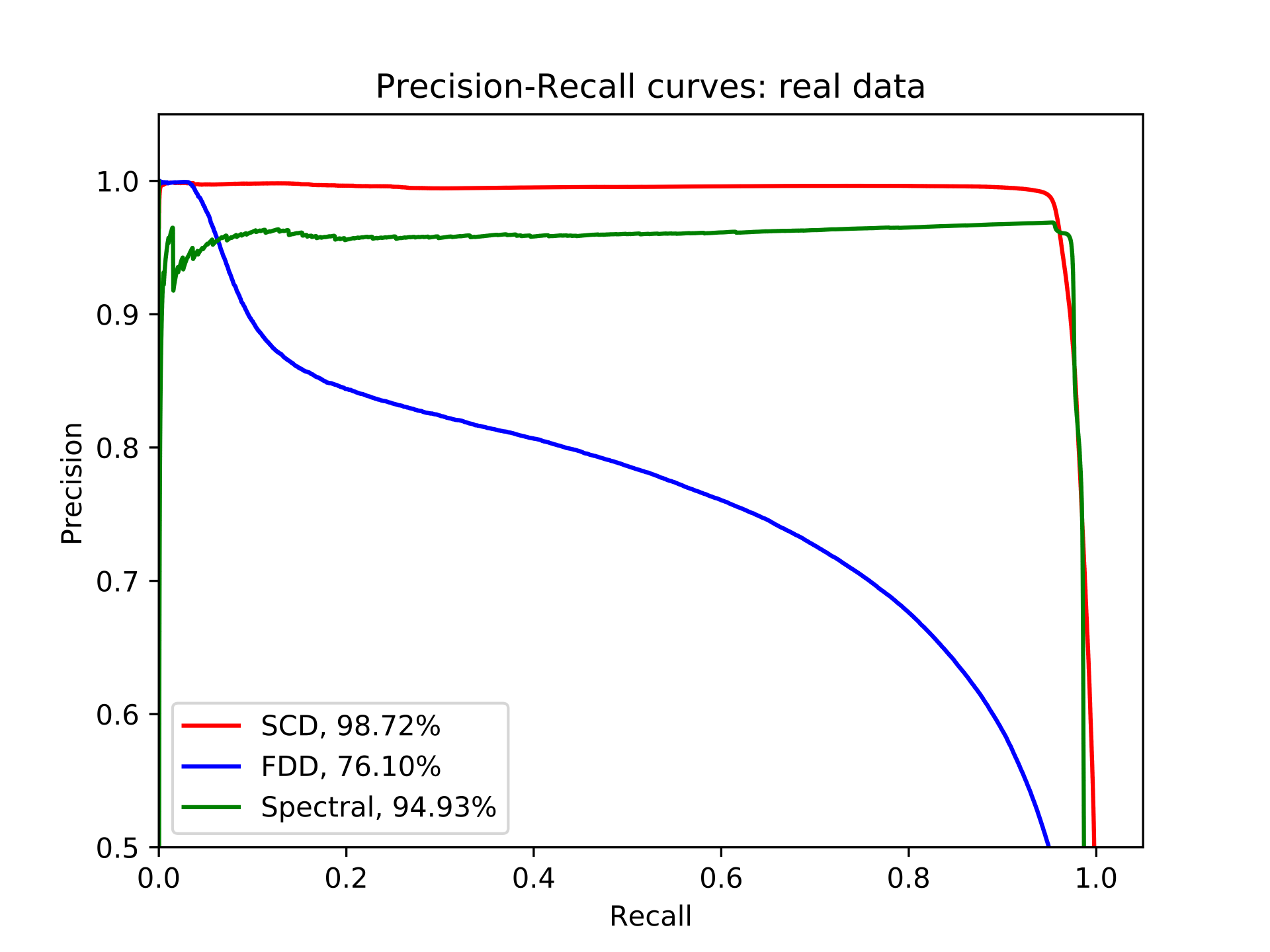}
	 \caption{{\bf Clustering real-world simplicial complexes.} The Precision-Recall curves achieved when using SCD to cluster real-world simplicial complexes.
 {\label{fig:scd_real2}}}
	\end{centering}
\end{figure}

We further validate our methodology by assessing its ability to correctly group our 1,775 real-world simplicial complexes.
We calculate the distances  between all pairs of the 1,775 real-world simplicial complexes, which results in distances between $1,775 \choose 2$ = 1,574,425 pairs for each of the three distance measures presented in Section \ref{sec:distances}.
As illustrated in Figure \ref{fig:scd_real1}, when the real-world simplicial complexes are embedded into 3D space based on their SCD distances by using multi-dimensional scaling, the simplicial complexes from the same data type group well together.
Indeed, the precision-recall curves presented in Figure \ref{fig:scd_real2} show that SCD achieves the highest classification performances (AP of 98.72\% and AUC of 99.58\%), followed by  spectral distance (AP of 94.93\% and AUC of 98.64\%) and by facet distribution distance (AP of 76.10\% and AUC of 93.11\%).

Taken together, these results demonstrate that SCD is a very sensitive measure of simplicial complex similarity.

\subsection{Uncovering biological information from PPI simplicial complexes}\label{sec:res_clustering}
In the experiments presented above, we measured the ability of simplets to capture global geometric features of simplicial complexes. In this section, we focus on the local geometry around nodes in simplicial complexes.
We assess if the local geometries of proteins in PPI simplicial complexes (which we capture with simplet degree vectors, see section \ref{sec:simplets}) relate to their functional annotations using two different methodologies: clustering and enrichment analysis of the resulting clusters, and canonical correlation analysis.
 
\subsubsection{Clustering and enrichment analysis}
In the first step, we investigate if proteins with similar local geometries (i.e., similar simplet degree vectors) tend to also have similar biological functions.
For both human and yeast, we computed the simplet degree similarity of the proteins in each of the two models of their interactomes (PPI network model and SC model, see section \ref{sec:models}).
We used these pairwise similarities as input for spectral clustering \cite{von2007}, which performs $k$-means clustering on the eigen-vectors of the matrix encoding the pairwise simplet degree similarities between the nodes.
Spectral clustering is favored over traditional $k$-means as it does not make strong assumptions on the shape of the clusters.
While $k$-means produces clusters corresponding to convex sets, spectral clustering can solve a more general problem such as intertwined spirals \cite{von2007}. 
To account for the randomness of the underlying $k$-means, each clustering experiments is repeated 10 times. For human and yeast simplicial complexes, we set the number of clusters, $k$, using the rule of thumb \cite{kodinariya2013}: $k = \sqrt{\frac{n}{2}}$, where $n$ is the number of nodes in the simplicial complex. I.e., we set  $k=90$ for human and $k=54$ for yeast data-sets. These values result in coherent clustering according to both sum of square error and normalized mutual information scores \cite{ana2003}. Then, we measure the biological coherence of the obtained clustering by the percentage of clusters that are statistically significantly enriched in at least one Gene Ontology (GO) annotation \cite{ashburner2000}. To this aim, we collected the experimentally validated GO annotations of genes from NCBI's entrez web portal (collected the 8$^{th}$ of March, 2018).
We considered GO biological process (GO-BP), GO molecular function (GO-MF), and GO cellular component (GO-CC) annotations separately. A cluster is statistically significantly enriched in a given annotation if the corresponding enrichment $p$-value is lower or equal to 5\% after Benjamini-Hochberg \cite{benjamini1995} correction for multiple hypothesis testing. To account for variability in cluster sizes, we also measure the biological coherence with the percentage of annotated genes that have at least one annotation enriched in their clusters.

\begin{figure}
	\begin{centering}
	\includegraphics[width=8cm]{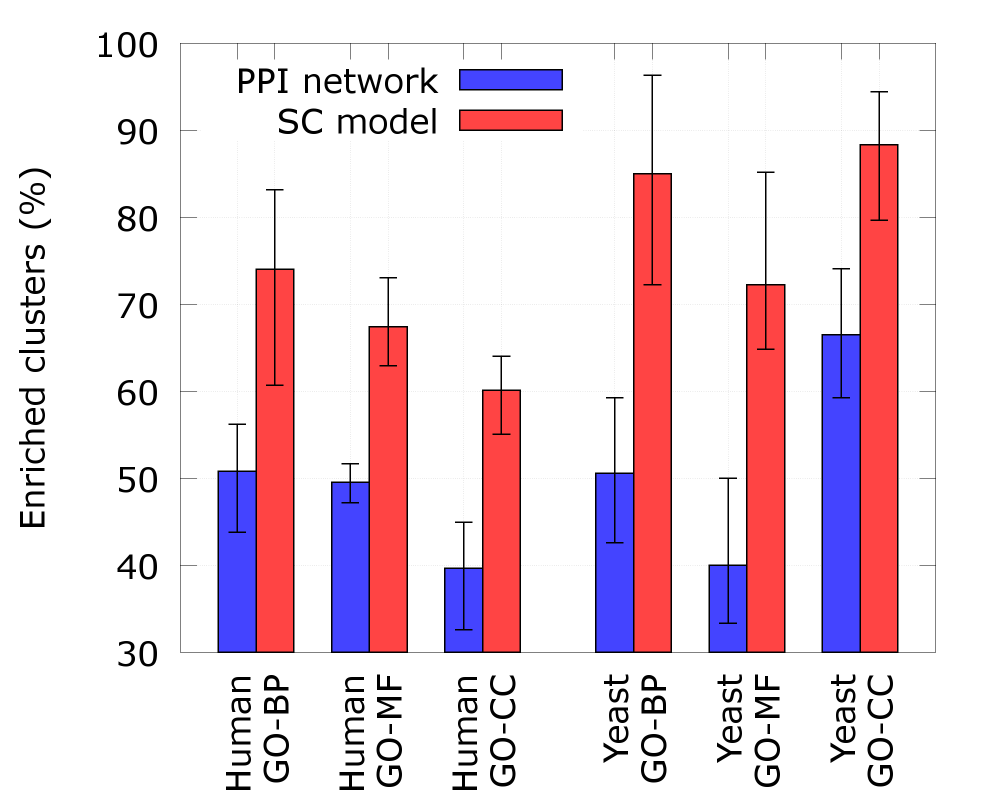}
	\caption{{\bf Biological relevance of clusters of genes,} as measured by the percentage of clusters having at least one enriched GO annotation. The error bars present minimum, average and maximum enrichment values over 10 runs of spectral clustering. 
	{\label{fig:clustering1}}}
	\end{centering}
\end{figure}

\begin{figure}
	\begin{centering}
	 \includegraphics[width=8cm]{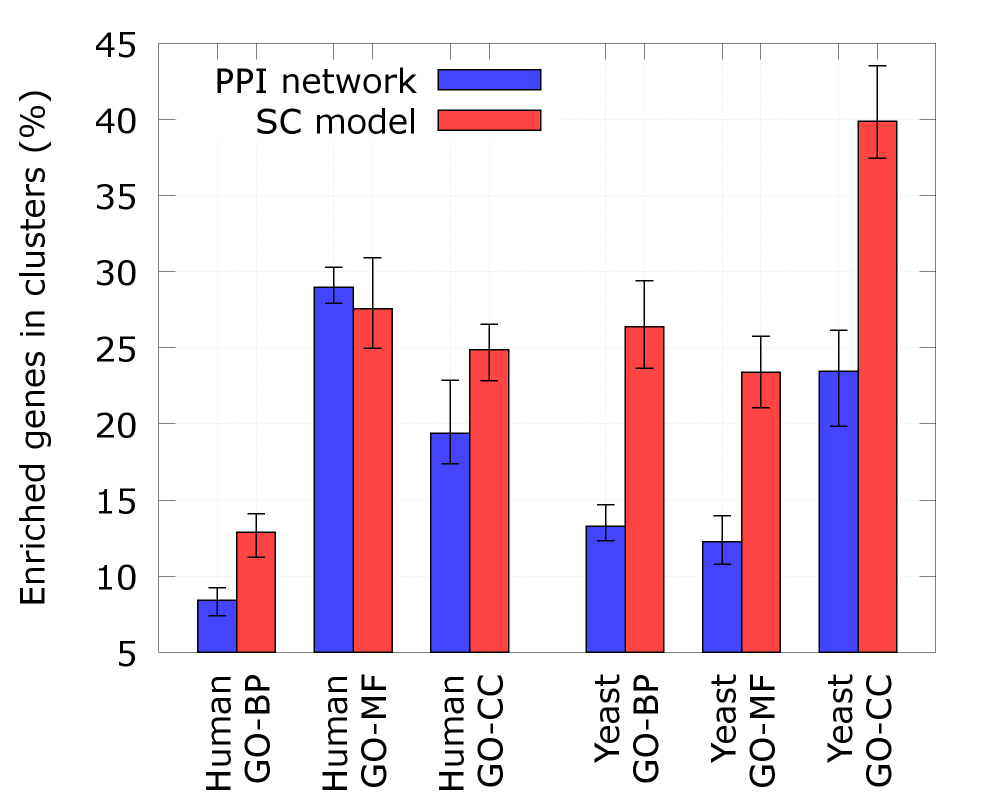}
	\caption{{\bf Biological relevance of clusters of genes,} as measured by the percentage of annotated genes having at least one GO annotation that is enriched in the cluster. The error bars present minimum, average and maximum enrichment values over 10 runs of spectral clustering. 
 {\label{fig:clustering2}}}
	\end{centering}
\end{figure}

As presented in Figures \ref{fig:clustering1} and \ref{fig:clustering2}, the clusters obtained from the SC models are more biologically coherent than the clusters obtained from the PPI network models.
Over all ten runs, for both species and for the three GO annotation types, the biological coherence in terms of enriched clusters is 50\% larger for the SC models than for the PPI network models, with 74.52\% of the clusters being enriched for the SC models and 49.51\% for the PPI network models.
In terms of the enriched genes in the clusters, the biological coherence is 46\% larger for the SC models than for the PPI network models, with 25.83\% of the genes being enriched for the SC models and 17.63\% for the PPI network models.

These results demonstrate that proteins having similar geometries in PPI networks modeled as simplicial complexes indeed have similar biological functions.
Using our simplets on the SC model allows for clusterings of proteins that best correspond to the hierarchical functional organization of the cell captured by GO biological process annotations, with about 81\% more of enriched proteins in the clusters obtained from the SC model by using simplets than on the PPI network by using graphlets (19.6\% for the SC model versus 10.8\% for the PPI network).
Similarly, the clusterings of proteins in the SC model best correspond to the organization of cell captured by GO cellular component annotations, with about 51.1\% more of enriched proteins in the clusters obtained from SC model than in the PPI network ones (32.4\% for the SC model versus 21.4\% for the PPI network).


For GO molecular function annotations, the results are slightly different. On average, the clusters of proteins on the SC models best group together proteins with similar molecular functions, with about 23.5\% more of proteins with enriched functions in the clusters obtained from the SC model than in the clustering obtained from the PPI network (25.5\% for the SC model versus 20.6\% for the PPI network).
But when considering on the human datasets alone, the PPI network achieves a slightly better performance (29.0\% for the PPI network versus 27.6\% for the SC model).
These lower performances on molecular functions are expected, because of the very good performances on biological processes. Recall that molecular functions capture the functions of proteins in isolation from each other, while biological processes correspond to higher-order biological functions that are performed by proteins collectively (in Gene Ontology, biological processes must involve more than one distinct molecular functions \footnote{\small \url{http://geneontology.org/page/ontology-documentation}}). Thus, maximizing the  enrichment of molecular function annotations and of biological process annotations are contradicting goals that cannot be expected to be optimized simultaneously.


\subsubsection{Canonical correlation analysis}

To further investigate the relationships between the local geometry around proteins in simplicial complexes (captured by their simplet degrees) and their biological functions (captured by their GO biological process annotations), we adapt the canonical correlation analysis (CCA) methodologies from \cite{yaveroglu2014}.
In our canonical correlation analysis (CCA) framework, the local geometry around $n$ proteins in a simplicial complex is captured in an  $n\times 32$ matrix, $R$, whose entry $R[v][i]$ is the $i^{th}$ simplet degree of node $v$.
Similarly, the biological functions of the proteins is captured in an  $n\times f$ matrix, $A$, whose entry $A[v][i]$ is 1 if protein $v$ is annotated by term $i$, and 0 otherwise. For both matrices, we excluded the genes that do not have any GO biological process annotations.

CCA is an iterative process that identifies linear relationships between the 32 simplet degrees and the $f$ GO biological process annotations.
First, CCA outputs two weight vectors, called {\em canonical variates}, so that the weighted sum of $R$ is maximally correlated with the weighted sum of $A$. The correlation between the two weighted sums is called their {\em canonical correlation}. After finding the first canonical variates, CCA iterates $min\left \{ 32, f \right \}$ times to find more weight vectors, such that the resulting canonical variates are not correlated with any of the previous canonical variates. We refer the interested reader to \cite{weenink2003} for the mathematical aspects of CCA.
\begin{figure}
	\begin{centering}
	 \includegraphics[width=8cm]{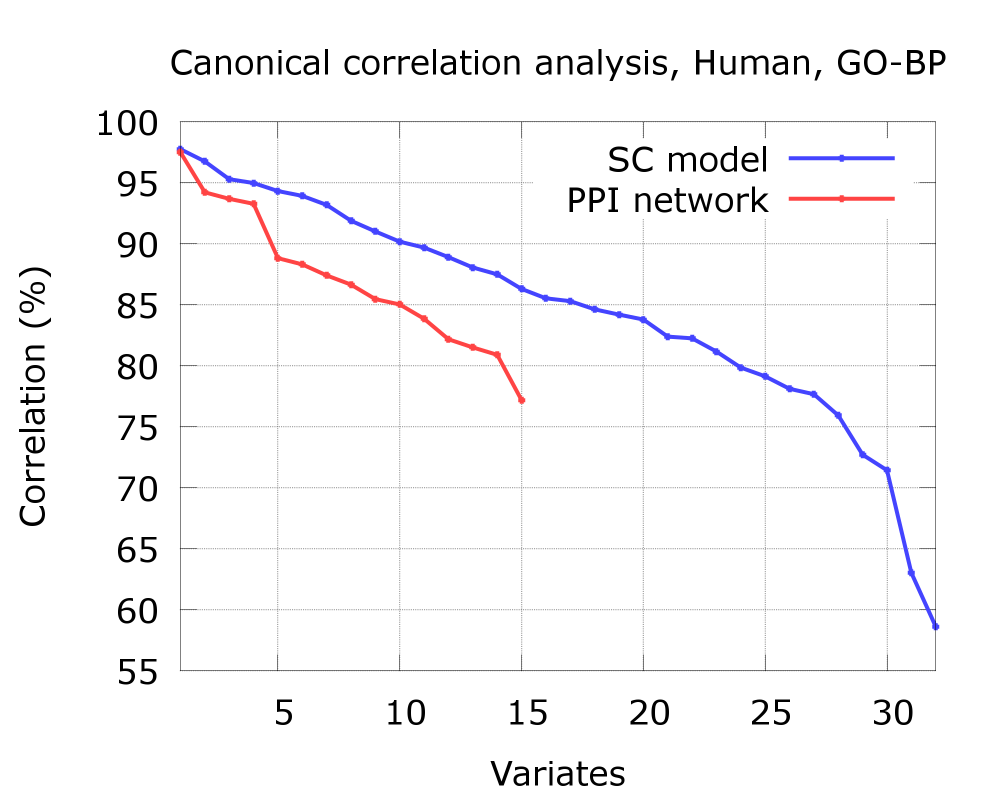}
	\caption{{\bf Canonical correlation analysis for human.} For a given simplicial complex, canonical correlation produces variates, which are linear combinations of go annotations and linear combinations of simplet degrees that best correlate over the nodes of the simplicial complex. For both models of human interactomes (PPI network and SC model), we plotted for each variate the corresponding correlation value (only statistically significantly correlated variates are presented, with canonical correlation $p$-value $\leq 5\%$). 
 {\label{fig:cca1}}}
	\end{centering}
\end{figure} 
\begin{figure}
	\begin{centering}
	 \includegraphics[width=8cm]{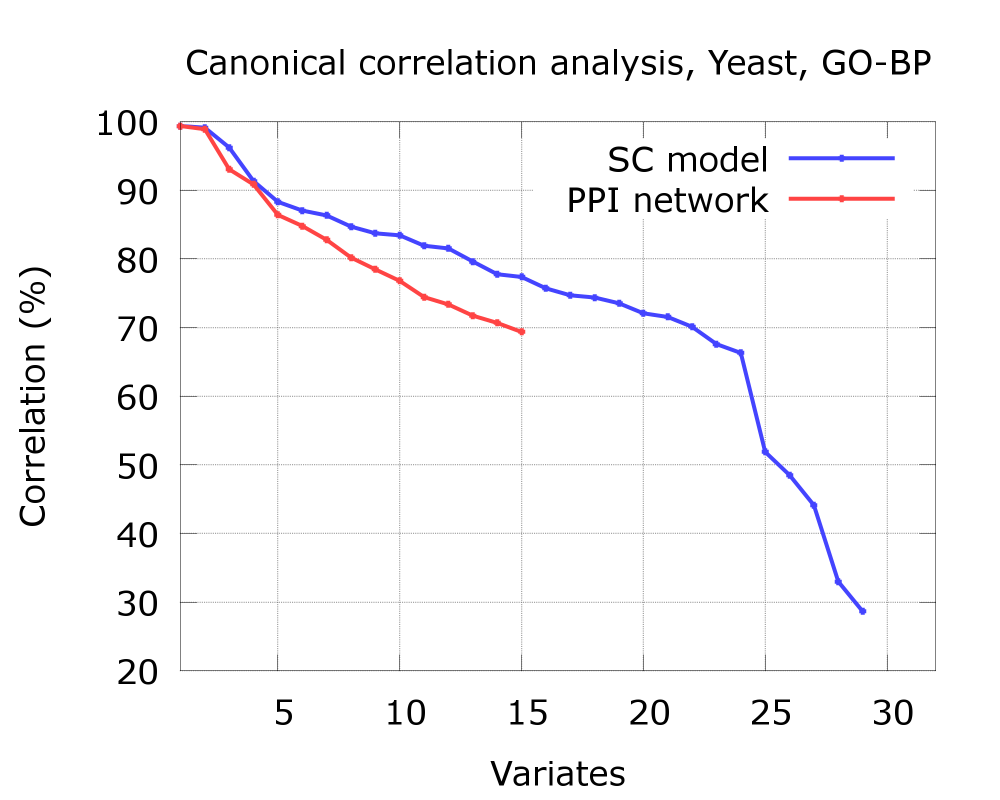}
	\caption{{\bf Canonical correlation analysis for yeast.} For both models of yeast interactomes (PPI network and SC model), we plotted for each variate the corresponding correlation value (only statistically significantly correlated variates are presented, with canonical correlation $p$-value $\leq 5\%$). 
 {\label{fig:cca2}}}
	\end{centering}
\end{figure} 
As presented in Figures \ref{fig:cca1} and \ref{fig:cca2}, the SC model allows for uncovering a larger number of linear relationships that the PPI network model. 
This is because only 15 out of the 32 simplets can appear in a 1-dimensional simplicial complexes, i.e., a PPI network, which correspond to the 15 2- to 4-node graphlets. Hence, CCA can only produce up-to 15 variates for a PPI network and up-to 32 variates for the SC model.
Moreover, these linear relationships have higher canonical correlations.
This means that by using simplets on the SC models we can capture more and better quality relationships between local geometry around nodes in simplicial complexes and their biological functions than if we use PPI networks. The same is observed when using GO cellular component and GO molecular function annotations (not shown due to space limitations).

\section{Conclusion}
We demonstrate that by the new way of accounting for multi-scale organization of PPI data both through modeling and new algorithms that we propose, we can uncover substantially more biological information than can be obtained by considering only pairwize interactions between proteins in PPI networks.
This pioneering observation can further be utilized to predict biological functions of unnanotated genes, which is a subject of further research.

We demonstrate the existence of the functional geometry in the PPI data by capturing the higher-order organization of these molecular networks by using simplicial complexes.
To mine the geometry of simplicial complexes, we propose simplets, which generalize graphlets to simplicial complexes. 
On randomly generated and real-world datasets, we define a sensitive measure of global geometrical similarity between simplicial complexes.
Also, we propose a higher-dimensional, simplicial complex-based (SC) model of a species' interactome, which combines protein-protein-interaction and protein complex data.
On human and yeast interactomes, by using clustering based on our new simplet-based measures of geometric similarity and cluster enrichment analysis, we show that our SC models are more biologically coherent than protein-protein interaction networks and that our simplets can efficiently mine this SC model as a new source of biological knowledge.
Furthermore, while we focus on simplicial complexes emerging from molecular network organization, our methodology is generic and can be applied to  multi-scale datasets from any scientific field, such as the multi-scale network data from physics, social sciences and economy.


\section*{Funding}
This work was supported by the European Research Council (ERC) Starting Independent Researcher Grant 278212, the European Research Council (ERC) Consolidator Grant 770827, the Serbian Ministry of Education and Science Project III44006, the Slovenian Research Agency project J1-8155 and the awards to establish the Farr Institute of Health Informatics Research, London, from the Medical Research Council, Arthritis Research UK, British Heart Foundation, Cancer Research UK, Chief Scientist Office, Economic and Social Research Council, Engineering and Physical Sciences Research Council, National Institute for Health Research, National Institute for Social Care and Health Research, and Wellcome Trust (grant MR/K006584/1) and UK Medical Research Council (MC\_U12266B).

\bibliographystyle{abbrv}
\bibliography{document}

\end{document}